\documentclass[12pt]{article}
\usepackage{epsfig}

\begin{document}

\title{Entanglement preservation in quantum cloning}

\author{Pawe{\l} Masiak\\
Institute of Physics, Polish Academy of Sciences, \\
Aleja Lotnik\'ow 32/46, 02-668 Warsaw, Poland}

\maketitle

\begin{abstract}
Recently Galv\~{a}o and Hardy 
have shown that quantum cloning can improve the performance of some quantum computation tasks. 
However such performance enhancement is possible only if quantum correlations survive the 
cloning process. 
We investigate preservation of the quantum correlations in the process of non--local 
cloning of entangled pairs of two--level systems. We consider different kinds of quantum cloning 
machines and compare their effectiveness in the cloning of non--maximally entangled pure 
states. A mean entanglement is introduced in order to obtain a quantitative evaluation of an average 
efficiency for the different cloning machines. 
We show that a reduction of the quantum correlations is significant and it strongly depends 
upon the kind of cloning machine used. Losses of the entanglement are largest in the case of 
the universal quantum cloning machine.
Generally, in all cases considered the losses of the entanglement are so drastic that the method 
of enhancement for the performance of the quantum computation using the quantum cloning seems 
to be questionable.
\end{abstract}

\section{Introduction}
Quantum correlations allow one to perform many communicational and computational 
tasks with efficiency unattainable using classical devices and they are at the heart of 
all considerations in quantum information theory. On the other hand the quantum 
correlations are very sensitive to any disturbances and easily suffer degradation. 
Any operation performed on a quantum correlated system can result in decreasing 
entanglement. 

One of the most important feature of information encoded in the quantum systems is that while 
classical informations can be copied exactly, quantum information cannot be cloned with perfect accuracy. 
More precisely, it is not possible to obtain a perfect copy of an arbitrary, unknown quantum state 
using unitary transformations. 
This so-called no-cloning theorem recognized by Wootters and Zurek \cite{Zurek}
follows directly from the linearity of quantum mechanics. 
A quantum cloning machine designed to perfect cloning of a given finite set 
of orthogonal states is used in a proof of this theorem.
The orthogonal states are cloned perfectly using this quantum cloning machine, 
but clones of all other states contain errors. 
This cloning machine is a special example of state-dependent cloners, which in general case 
produce approximate clones of the chosen non-orthogonal states 
\cite{Buzek_asym,Bruss_asym,Barnett_asym}. 
State--independent cloning machines producing clones, whose quality is independent of 
input states, form another class of the cloners. Bu\v{z}ek and Hillary first showed how the 
no--cloning theorem can be bypassed \cite{Buzek_beyond}. 
They also proposed construction of a network, consisting of quantum gates, realizing 
the idea of an universal cloning machine \cite{Buzek_network}.   
This network clones any quantum state of a qubit equally well, where the figure of merit 
is the local fidelity, i.e. an overlap between the states of the input and the output qubits. 
These results were extended later to cases where $d$--dimensional systems, for $d>2$, are cloned 
and to cases where more than two clones are obtained in the cloning process \cite{Werner_d1,Buzek_d,Werner_d2}. 

Quantum cloning is considered in quantum cryptography as a useful tool allowing 
a partial break in security of some quantum cryptography protocols \cite{Bruss_asym}. 
Recently Galv\~{a}o and Hardy have shown that quantum cloning can be also applied 
to improve the performance of some quantum computation tasks \cite{Hardy_qcomp}. 
Preservation of the quantum correlations is specially important in this case because 
the computational power of quantum computers is based on the use of highly correlated quantum 
systems. Cloning can be considered as a method of enhancement of the efficiency of the quantum 
computers only if it allows one to produce clones which are not only similar to 
the cloned systems, but in the first place, if it preserve the entanglement of the states.
 
The article has the following structure. We begin with short characterization of the 
three kinds of deterministic cloning machines. 
Next, we recall a definition of the entanglement of formation, which is used 
to measure the quantum correlations and finally, in the main part of the article, 
we present results of calculations of the entanglement of formation for the clones 
obtained using the cloning machines considered and we compare the efficiency 
of these cloners with regards to the preservation of the quantum correlations.

\section{Quantum cloning machines}

First we consider the Wootters-Zurek cloning machine. Originally Wootters and Zurek considered cloning
of qubits, i.e. states belonging to a two-dimensional Hilbert space. We want to study the cloning of  
the correlated systems, so we need to extend the Wootters-Zurek results. We propose an extended 
version of the Wootters-Zurek cloning machine adapted to a task of cloning the entangled states of 
the two qubit systems. We choose a basis of Bell states as the set of states which are perfectly cloned: 
$|\varphi_{(1,2)}\rangle=|\Phi^{\pm}\rangle=\frac{1}{\sqrt{2}}(|00\rangle \pm |11\rangle)$ and 
$|\varphi_{(3,4)}\rangle=|\Psi^{\pm}\rangle=\frac{1}{\sqrt{2}}(|01\rangle \pm |10\rangle)$. 
The cloning process is defined by a transformation relation on the states $|\varphi_i\rangle$:
\begin{equation}
|\varphi_i \rangle_1 |\hbox{\hspace{0.25cm}}\rangle_2 |\omega\rangle_C \longrightarrow
|\varphi_i \rangle_1 |\varphi_i \rangle_2|\omega_i\rangle_C.
\end{equation}
At the beginning of the cloning process the second pair of qubits is in a given, but 
non-specified blanc state $|\hbox{\hspace{0.25cm}}\rangle_2$ and the cloning machine 
is in an initial state $|\omega\rangle_C$. After the transformation both pairs are in 
the cloned state $|\varphi_i \rangle$, and the cloning machine is in the final 
state $|\omega_i\rangle_C$ depending upon the initial state of the input pair of the qubits. 

We consider not too much restrictive assumption that the states $|\omega_i\rangle$ of 
the cloner are orthogonal, i.e. $\langle\omega_i|\omega_j\rangle=\delta_{ij}$.
An arbitrary pure superposition of the basis states 
$|\psi_{in}\rangle=\sum_{i=1}^4\alpha_i|\varphi_i\rangle$ transforms as follows:
\begin{equation}
\sum_{i=1}^4\alpha_i|\varphi_i\rangle_1 |\hbox{\hspace{0.25cm}}\rangle_2 |\omega\rangle_C 
\longrightarrow
\sum_{i=1}^4\alpha_i|\varphi_i\rangle_1 |\varphi_i\rangle_2 |\omega_i\rangle_C
=|\Psi\rangle^{(out)}_{12C}.
\end{equation}
A reduced density operator of any clone reads:
\begin{eqnarray}
\hat{\rho}^{\,(out)}_{k=1,2}
& = & Tr_{k=2,1}\left[Tr_C\left[|\Psi\rangle^{(out)}_{12C}
   {}^{(out)}_{12C}\langle\Psi|\right]\right]\\ \nonumber
& = & \sum^4_{i=1}\alpha^2_i|\varphi_i\rangle \langle\varphi_i|,
\end{eqnarray}
where for simplicity we have taken the probability amplitudes, $\alpha_i$, real. 
The fidelity of the clones $F(\alpha_i)
=\langle\psi_{in}| \hat{\rho}^{\,(out)}_k |\psi_{in}\rangle=\sum^4_{i=1}\alpha^4_i$.
We see that in agreement with the scheme of this cloning machine, the basis states 
are cloned perfectly and the clones of all other states contain some errors. 
This is illustrated by changes of the fidelity, which is less than one for these states. 
The minimal fidelity $F_{min}=\frac{1}{4}$ and it is less than the fidelity attained in 
a procedure for the estimation of states using an optimal measurement, 
which can be regarded as a classical counterpart of the quantum cloning 
\cite{Buzek_class}. 
The mean fidelity of such classical cloning $\overline{F}_{class}=\frac{2}{5}$.
Eight product states of a form 
$|\psi_{in}\rangle=\sum^4_{i=1} \alpha_i|\varphi_i\rangle$, where $\alpha_i=\pm\frac{1}{2}$,
are cloned the worst. These states are fortunately least interesting from our point of view. 
The Wootters-Zurek cloning machine clones some states very badly. However it has one big 
advantage: the quantum correlation of some kind of states are completely preserved in the cloning.

Next under consideration is the best known and most widely studied, the state-independent 
symmetric optimal quantum cloning machine \cite{Buzek_beyond,Buzek_network}. 
The cloning transformation is a shrinking type transformation, 
which can be formally expressed in the form: 
$\hat{\rho}^{\,(out)} \longrightarrow s\, \hat{\rho}^{\,(id)}+\frac{1-s}{4}\, \hat{\bf 1 }$, 
where the shrinking factor $s=\frac{M+4}{5\,M}$, $\hat{\rho}^{\,(id)}$ is a density matrix 
of the ideal clone of the input state and $M$ is the number of clones produced \cite{Werner_d1,Werner_d2}. 
All clones are in this same final state described by the density matrix $\hat{\rho}^{\,(out)}$.
We restrict our investigations to a case when only two clones are produced due to, first, it 
being the generic case, and second, for more clones the entanglement of each of them rapidly 
decreases with the number of clones \cite{Masiak}. It is not possible to obtain arbitrarily many entangled 
clones of the input state even if the cloned pair of the qubits is at the beginning of the cloning 
process in the maximally entangled state. The clones are entangled if at most five copies 
are produced. More copies will already be in states with no quantum correlation \cite{Masiak}.

At the end of the survey on the classes of the deterministic quantum
cloning machines we turn our attention on the class of asymmetric
state-independent cloning machines \cite{Ceft}. 
In this case the states of the clones are different. They are described by density matrices 
obtained as a result of action similar to before, i.e. shrinking type transformations. 
However now the scaling factors can be different: 
\begin{equation}
\hat{\rho}^{\,(out)}_i \longrightarrow s_i\, \hat{\rho}^{\,(id)}+\frac{1-s_i}{4}\, \hat{\bf 1 },
\end{equation}
where, as previously, we have restricted the considerations to the case of two clones, i.e. 
$i=1,2$. The parameters $s_i$ are connected by a relation:
\begin{equation}
4(1-s_1-s_2)^2-(1-s_1)(1-s_2)\leq 0,
\end{equation}
which define a range of allowed values of them. 
In extremal cases the asymmetric cloning transformation degenerates: when $s_1=1$ and $s_2=0$ 
the states of the qubits do not change at all, and when $s_1=0$ and $s_2=1$ the states of 
the qubits swap; and actually these transformations should not be called the cloning.
When both shrinking parameters are equal to $\frac{3}{5}$ the symmetric cloning machine 
is reproduced. 

\section{Entanglement of formation of clones}

The non-locality of the quantum state can manifest itself in many different ways. 
The best known and one which has been tested in many experiments is 
a violation of the Bell inequality \cite{Aspect1,Aspect2}. 
However, the violation of the Bell inequality is not wholly satisfactory measure of 
the quantum correlations. There are other quantities which have been 
developed as measures of the quantum non-locality 
\cite{Vedral_ent1,Vedral_ent2,Wootters1,Wootters2}. 
In the case of a pair of correlated qubits, one such measure is the entanglement 
of formation \cite{Wootters1,Wootters2}. 
This quantity is much more sensitive to the degree of quantum correlations than 
the Bell inequality and what is also very important is that a finite, compact analytical 
formula for it is known. 
The entanglement of formation $E[\hat{\rho}]$ is defined in the following way 
\cite{Wootters1,Wootters2}:
\begin{equation}
  E[\hat{\rho}]={\mathcal E}(C(\hat{\rho})),
\end{equation}
where
\begin{equation}
  {\mathcal E}(y)=h\left( \frac{1+\sqrt{1-y^{2}}}{2}\right) ,
\end{equation}
$h(x)=-x\log _{2}(x)-(1-x)\log _{2}(1-x)$ and $C(\hat{\rho})$ is so-called concurrence 
of the two-qubit state $\hat{\rho}$ given by expression:
\begin{equation}
  C(\hat{\rho})=\mbox{max}\,\{0,\lambda _{1}-\lambda _{2}-\lambda _{3}-\lambda_{4}\}.
\end{equation}
The $\lambda _{i}$'s in these
expressions are the square roots of eigenvalues, in decreasing order, of a
non-Hermitian matrix $\hat{\rho} \hat{\tilde{\rho}}$, and $\hat{\tilde{\rho}} 
=(\hat{\sigma}_{y}\otimes \hat{\sigma}_{y})\hat{\rho}^{*}(\hat{\sigma} _{y}
\otimes \hat{\sigma}_{y}).$ 
We use the entanglement of formation to investigate preservation of the quantum 
correlations in the cloning process. We compare the three above-described  
cloning machines used to produce the clones of the non-local system consisting 
of two qubits formed in the entangled pure state. 
Any of the Bell states can be chosen:
\begin{eqnarray}
|\Phi ^{\pm }(\alpha )\rangle &=& \alpha |00\rangle \pm \beta |11\rangle, \nonumber \\
|\Psi ^{\pm }(\alpha )\rangle &=& \alpha |01\rangle \pm \beta |10\rangle, \nonumber 
\end{eqnarray}
where $\alpha$ determines the amount of entanglement in the state, $\beta =\sqrt{1-\alpha ^{2}}$
and for simplicity both $\alpha$ and $\beta$ are taken as real.
We decide to use the state $|\Psi^{-}(\alpha)\rangle$.
The states of the clones at the output of the cloning machines considered are described by 
the following density matrices.
\begin{enumerate}
\item Wootters-Zurek cloning machine (WZCM)
\begin{equation}
\hat{\rho}^{\,(out)}_{WZCM}(\alpha)
  =\frac{1}{2}[|01\rangle \langle 01|+|10\rangle \langle 10|
\label{WZCM} 
\end{equation}
$$
   -2 \alpha \beta (|01\rangle \langle 10| +|10\rangle \langle 01|)].
$$
\item Symmetric universal cloning machine (SCM) \cite{Buzek_beyond}
\begin{equation}
\hat{\rho}^{\,(out)}_{SCM}(\alpha)=\frac{6\,\alpha^2+1}{10}\,|01\rangle \langle 01|
    +\frac{6\,\beta^2+1}{10}\,|10\rangle \langle 10|
\label{SCM} 
\end{equation}
$$
    -\frac{3\,\alpha \beta }{5}\,(|01\rangle \langle 10| +|10\rangle \langle 01|)
    +\frac{1}{10}\,(|00\rangle \langle 00|+|11\rangle \langle 11|).
$$
\item Asymmetric universal cloning machine (ACM) \cite{Ceft}
\begin{equation}
\hat{\rho}^{\,(out)}_{i,\,ACM}(\alpha|s_i)=\frac{(4\,\alpha^2-1)s_i+1}{4}\,|01\rangle \langle 01|
    +\frac{(4\,\beta^2-1)s_i+1}{4}\,|10\rangle \langle 10|
\label{ACM} 
\end{equation}
$$
    -s_i\,\alpha \beta \,(|01\rangle \langle 10| +|10\rangle \langle 01|)
    +\frac{1-s_i}{4}\,(|00\rangle \langle 00|+|11\rangle \langle 11|).
$$
\end{enumerate}

\begin{figure}[t]
\begin{center}
\centerline{\includegraphics[width=\columnwidth,clip]{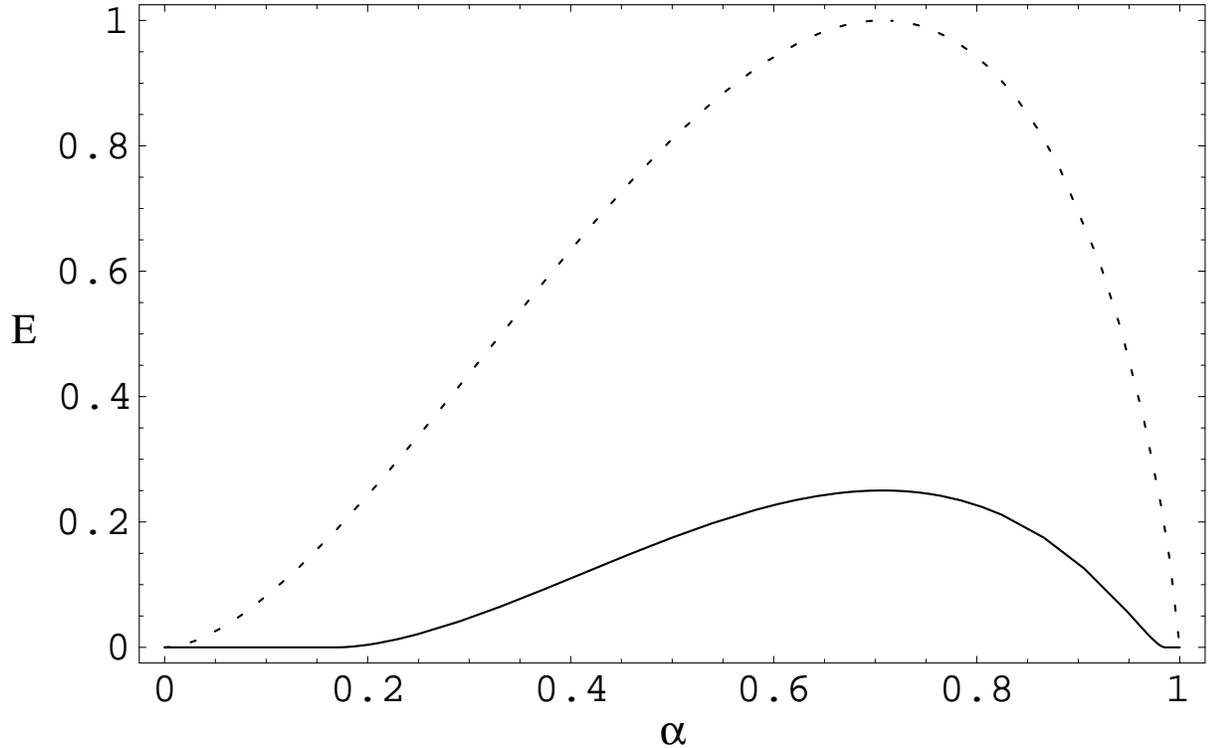}}
\caption{Entanglement of formation of the clones obtained using WZCM (dashed curve) 
and SCM (solid curve) as a function of the parameter $\alpha$.}
\label{fig1}
\end{center}
\end{figure}

\begin{figure}[t]
\begin{center}
\centerline{\includegraphics[width=\columnwidth,clip]{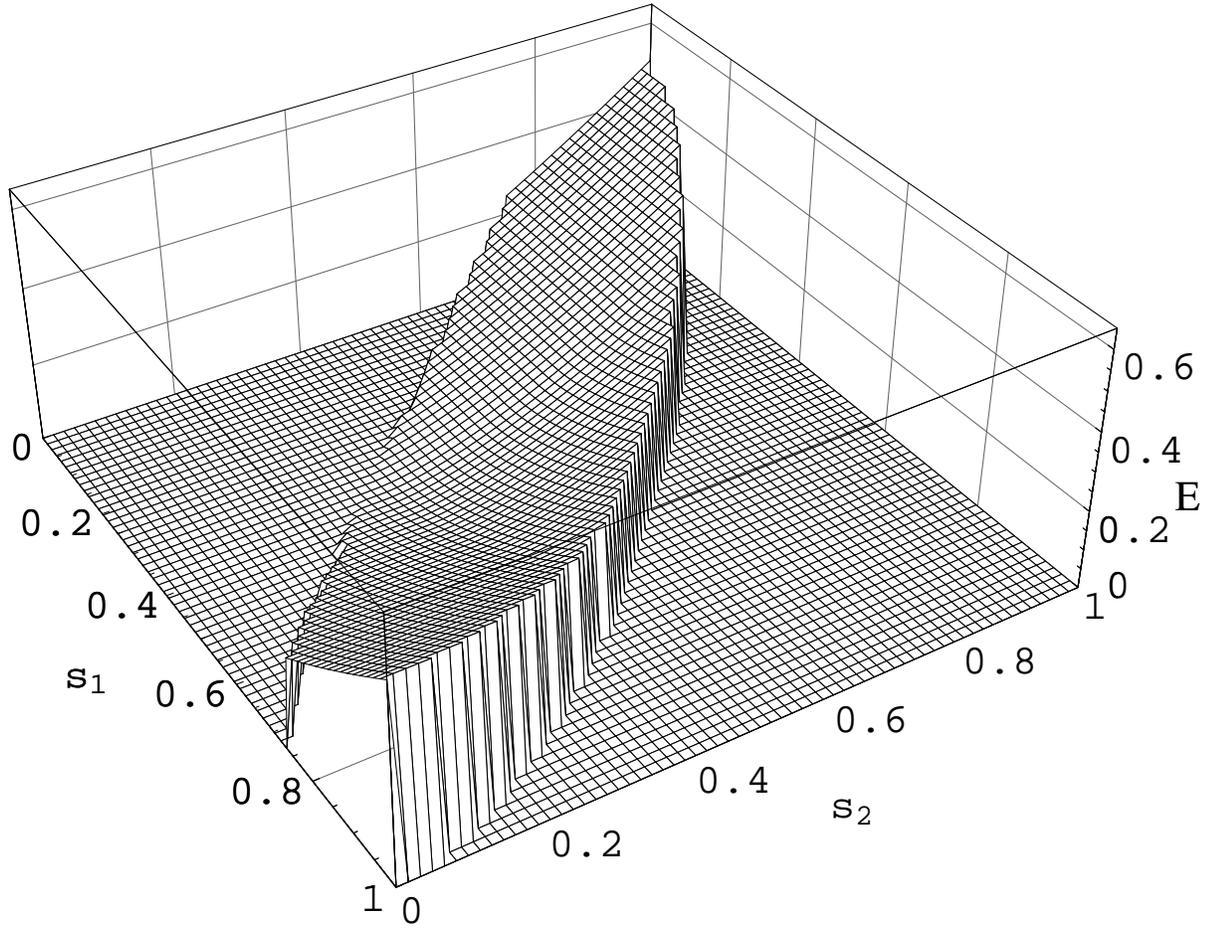}}
\caption{Average entanglement of formation of the clones produced 
from the singlet state $|\Psi^{-}\rangle$ using ACM 
as a function of the shrinking parameters $s_1$ and $s_2$.}
\label{fig2}
\end{center}
\end{figure}

\begin{figure}[t]
\begin{center}
\centerline{\includegraphics[width=\columnwidth,clip]{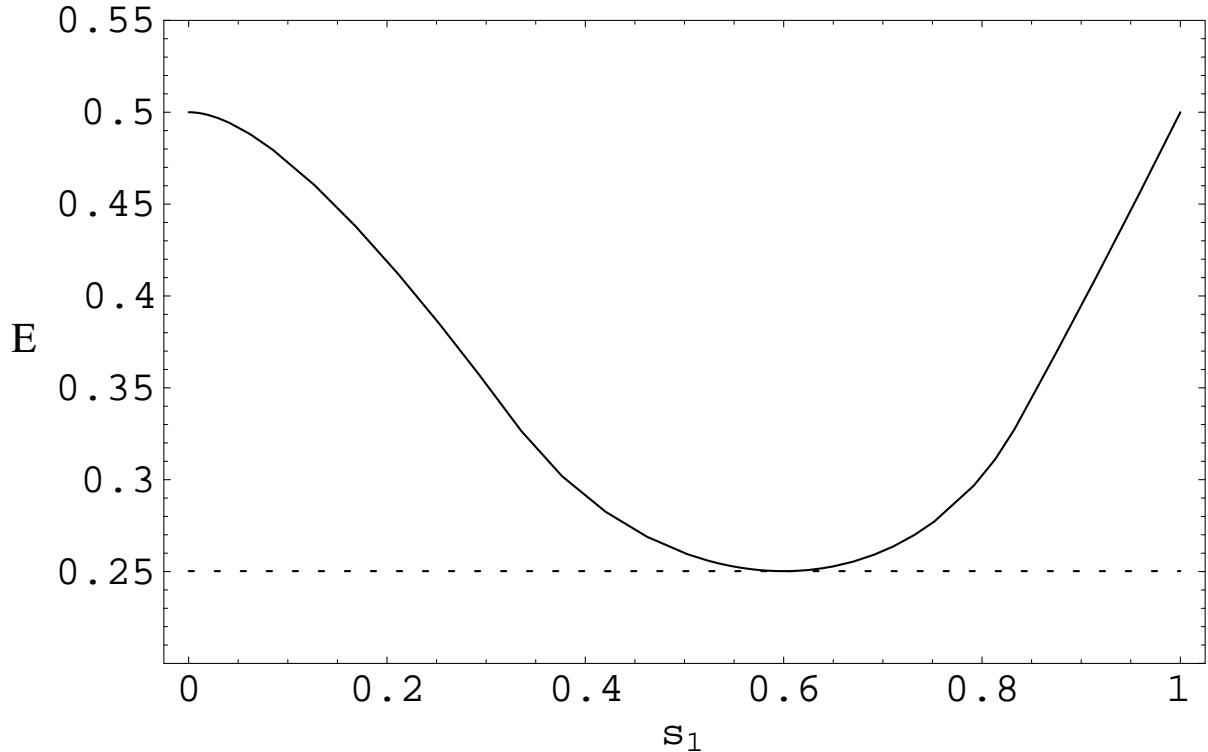}}
\caption{Average entanglement of formation of the clones produced from 
the singlet state $|\Psi^{-}\rangle$ using ACM for the shrinking parameters $s_1$ 
and $s^{+}_2(s_1)$ as a function of the parameter $s_1$. The dashed line shows 
the entanglement of formation of the clones obtained using SCM.}
\label{fig3}
\end{center}
\end{figure}

\begin{figure}[t]
\begin{center}
\centerline{\includegraphics[width=\columnwidth,clip]{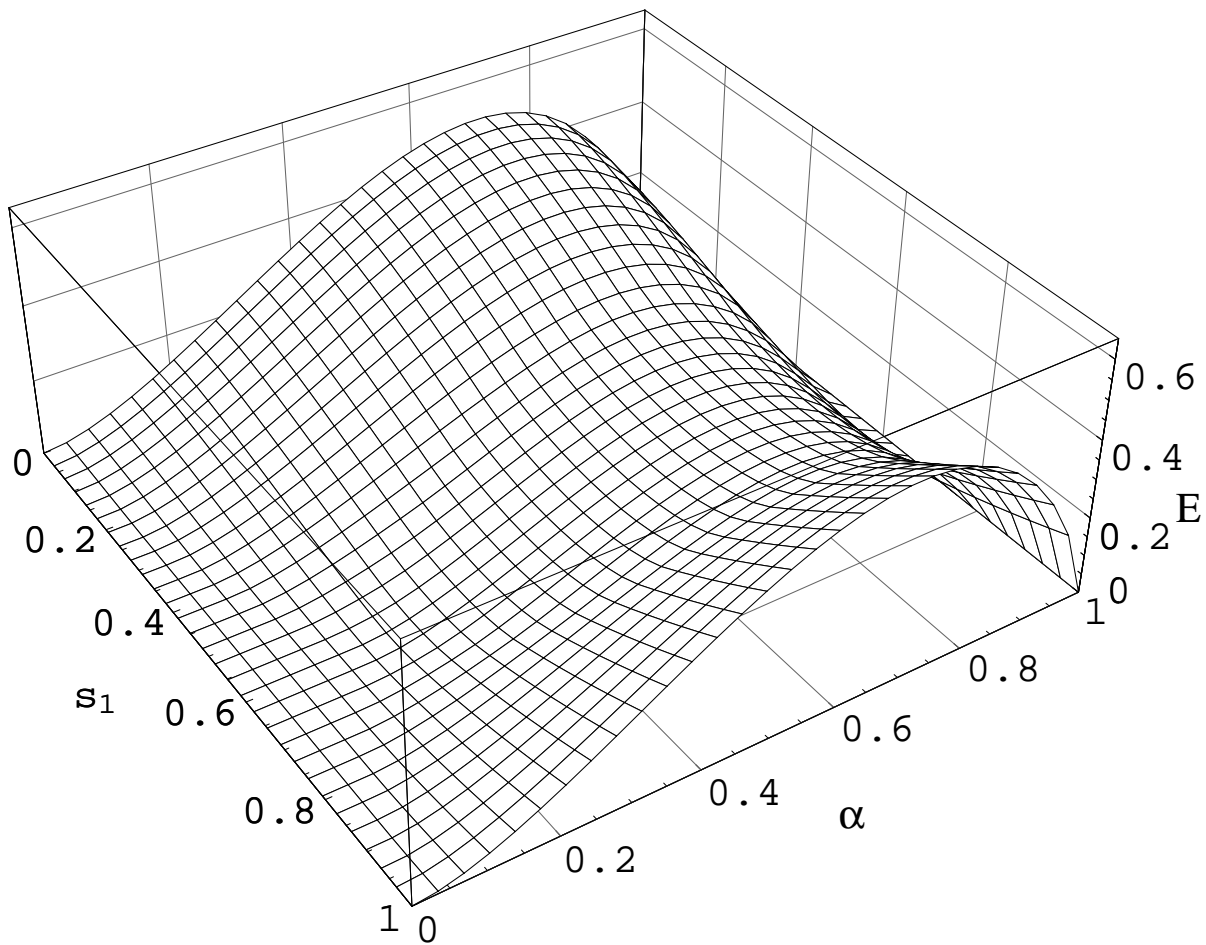}}
\caption{Average entanglement of formation of the clones of the pure states 
$|\Psi ^{-}(\alpha)\rangle$ produced using ACM calculated for scaling parameters 
$(s_1,s^{+}_2(s_1))$ as a function of the parameters $\alpha$ and $s_1$.}
\label{fig4}
\end{center}
\end{figure}

We calculate the entanglement of formation of the output states of the cloning machines 
described above. In figure \ref{fig1} we show the entanglement of formation of the clones 
of the states $|\Psi ^{-}(\alpha)\rangle$ obtained as the result of cloning using the 
symmetric universal cloning machine and the Wootters-Zurek cloning machine. 
The clones produced by the SCM distinctly lose the entanglement, not only the value 
of entanglement for them is strongly reduced but also some states, which, at 
the beginning had less entanglement, but have now lost all of it. 
The degree of reduction depends upon how much entanglement the state had before the cloning 
and is maximally equal to a factor $\frac{1}{4}$ for the singlet state 
$|\Psi ^{-}\rangle=|\Psi ^{-}(\frac{1}{\sqrt{2}})\rangle$. 
The entanglement of the clones obtained in cloning by the WZCM is equal to the entanglement 
of the initial states which were cloned. 
The prize for entanglement preservation, which must be paid in this case, 
is dependence of the fidelity upon the input state of the cloning transformation.
An analysis of the asymmetric cloning machine is more complex because two additional 
parameters characterizing the cloning transformation are presented in this case. 
In order to describe a mean amount of the entanglement falling to the clones we define 
an average entanglement per clone:
\begin{equation}
E^{\,(av)}_{ACM}\,(\alpha|s_1,s_2)
=\frac{1}{2}\sum_{i=1,2} E[\hat{\rho}^{\,(out)}_{i,ACM}(\alpha|s_i)].
\end{equation}
In figure \ref{fig2} we show the average entanglement 
$E^{\,(av)}_{ACM}\,(\frac{1}{\sqrt{2}}|s_1,s_2)$ 
of the clones of the singlet state $|\Psi^{-}\rangle$ for values of the scaling 
parameters $s_1$ and $s_2$ belonging to the region given by Eq. (5). 
The maximum of the average entanglement is attained for two 'degenerate' cases, 
when the state of the cloned pair of qubits does not change or when the cloned pair 
and the blanc pair swap states. We exclude these two cases from our analysis 
because cloning actually do not happen in them.
We see that the average entanglement is largest for values $s_1$ and $s_2$, which lie
on a curve bordering the region containing allowed values of the parameters $s_i$
from a side of larger values of $E^{\,(av)}_{ACM}$, the curve described by the equation: 
$s^{+}_2(s_1)=\frac{1}{8}(7-7s_1+\sqrt{1+14s_1-15s_1^2})$ and attains minimum on 
the curve which borders the region considered from the opposite side, the curve given by the equation 
$s^{-}_2(s_1)=\frac{1}{8}(7-7s_1-\sqrt{1+14s_1-15s_1^2})$. 
We concentrate our attention on these values of the parameters $s_i$, which correspond 
to largest values of the average entanglement. 
In figure \ref{fig3} the average entanglement 
$E^{\,(av)}_{ACM}\,(\frac{1}{\sqrt{2}}|s_1,s_2)$ 
of the clones of the singlet state for parameters $s_1$ and $s_2=s^{+}_2(s_1)$
is presented. A dashed line corresponding to the value of the entanglement of the clones 
obtained using the symmetric cloning machine is used as a reference. 
It turn out that the asymmetric cloning machine can, on average, be evidently better 
at cloning the entangled states than the symmetric cloning machine, which produces identical 
but less entangled clones. The ACM allows one to preserve in both clones more entanglement 
at the cost of a non-uniform distribution of its between the produced clones.
The minimal value of the average entanglement is obtained for $s_1=\frac{3}{5}=s_2$, which 
is precisely the value of the shrinking parameter of the SCM. 
So far we have studied the average entanglement of the clones of the singlet state. 
It is the most representative quantity because the singlet state is maximally entangled state.
To complete the considerations concerning the asymmetric cloning machine 
we show in figure \ref{fig4} the average entanglement $E^{\,(av)}_{ACM}\,(\alpha|s_1,s^{+}_2(s_1))$ 
for the clones of states $|\Psi^{-}(\alpha)\rangle$ for the scaling parameters 
$(s_1,s^{+}_2(s_1))$ lying on the curve corresponding to maximal values of 
the average entanglement.
We see that also for states other than the singlet state the ACM is more efficient in preserving 
of the entanglement. The maximum of the average entanglement 
is obviously attained for the singlet state and for this state the dependence upon the shrinking 
parameters is most evident. For other input states the average entanglement is already less 
sensitive to changes of the parameter $s_1$.

\begin{figure}[t]
\begin{center}
\centerline{\includegraphics[width=\columnwidth,clip]{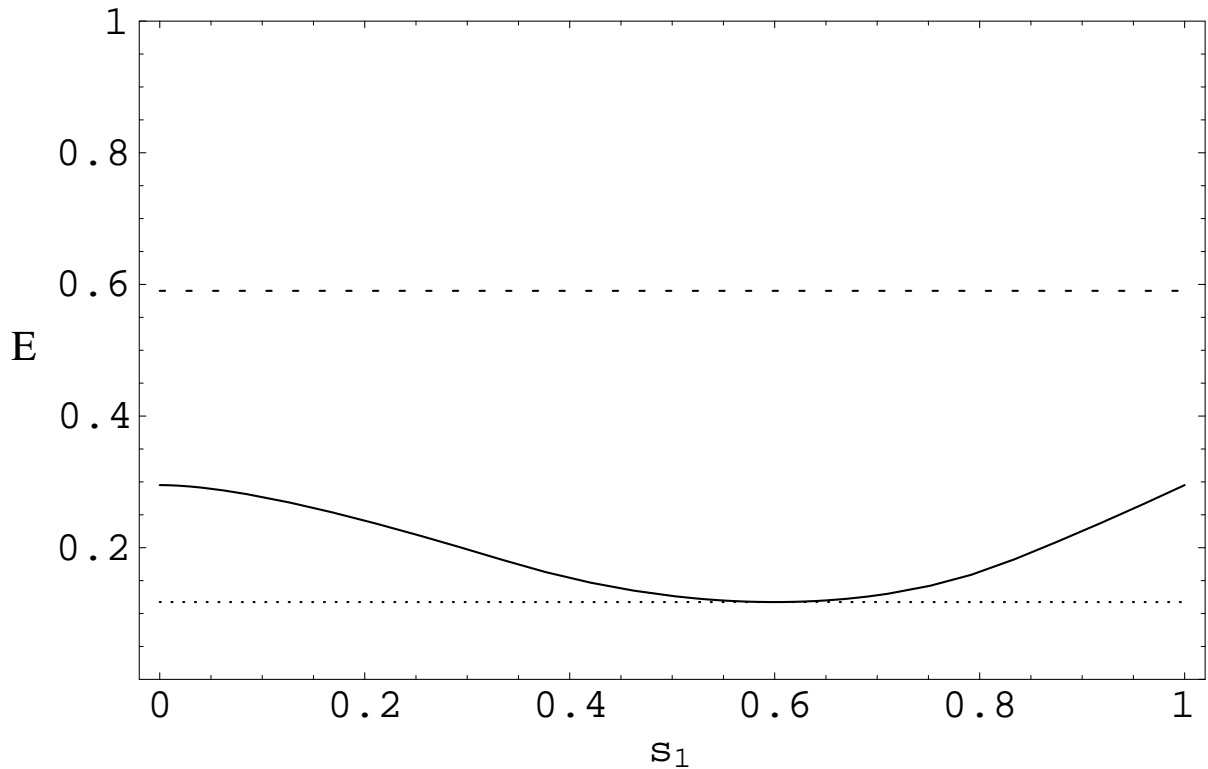}}
\caption{Mean entanglement of formation of clones produced using ACM (solid curve), 
using WZCM (dashed line), and SCM (dotted line) as a function of the parameters $s_1$.}
\label{fig5}
\end{center}
\end{figure}

Comparison of the cloning machines using the entanglement of formation calculated for 
the clones of the states $|\Psi^{-}(\alpha)\rangle$, presented above, informs us how many 
of the quantum correlations survive the cloning when the given state is cloned. 
We would like also to know, which of the considered cloning machines is better, more efficient, 
preserve more entanglement on average, when the cloned states are chosen arbitrarily. 
In order to analyse quantitatively such a quantity we define a new parameter: a mean entanglement 
of a family of the states, calculated as the average of the entanglement of formation over the whole 
ensemble of the input states:
\begin{eqnarray}
\overline{E}_b
& = & \int^{1}_{0} E[\hat{\rho}^{\,(out)}_{b}(\alpha)]\, d\alpha, \\
\overline{E}_{ACM}(s_1,s_2) & = & \int^{1}_{0} E^{\,(av)}_{ACM}(\alpha|s_1,s_2)\, d\alpha,
\end{eqnarray}
where an index $b=WZCM$, $SCM$, specifies the kind of cloning transformation.
In the case of the asymmetric cloning machine the mean entanglement still is a function of 
the scaling parameters $s_i$. In the both remaining cases the mean entanglement is just 
a number characterizing the global ability of these cloning machines to preserve 
the quantum correlations in the cloning process. 
The mean entanglement of formation for the WZCM and the SCM are, respectively, 
$\overline{E}_{WZCM}=0.59026$ and $\overline{E}_{SCM}=0.11747$.
In figure \ref{fig5} we show the mean entanglement of the clones obtained using 
the ACM for the shrinking parameters $(s_1,s^{+}_2(s_1))$.
We see that the mean entanglement of formation is larger when the ACM produces more asymmetric 
clones and tends to a minimal value for the scaling parameters corresponding to the SCM.
The mean entanglement is largest for the WZCM, the result of the cloning process using the ACM 
is much poorer and the mean entanglement of the clones obtained using the SCM is already very small 
and close to zero. 

\section{Conclusions}
To summarize, in this article we have investigated how the entanglement of formation 
of the pairs of qubits changes in the cloning process. We have analysed three kinds of the cloners: 
the modified version of the state-dependent Wootters-Zurek cloning machine designed to 
produce perfect clones of the Bell states, the universal symmetric cloning machine producing 
similar clones and the state-independent asymmetric cloning machine producing clones which are 
different. We have computed and compared the entanglement of formation of the clones of 
the entangled pure states. We have shown that the entanglement degrades very distinctly 
in the case of the SCM. The clones produced using the ACM are much more entangled, 
however the entanglement is non-uniformly distributed among the clones. Also, more entanglement 
is preserved on average in the case of the ACM, when mean value of the entanglement per clone 
is analysed. In the case of the WZCM the whole entanglement of the cloned systems survive. 
We have also defined the new quantity, the mean entanglement calculated as the mean over 
the whole ensemble of the cloned states, the quantity, which characterizes the cloning machines 
globally. The mean entanglement seems to be large enough for the WZCM to hope 
that such a cloner can be useful in practical applications, but the fidelity of the clones produced
is rather poor in this case. In both remaining cases the mean entanglement is much smaller. 

We conclude that the quantum cloning machines can be used to produce the clones of entangled 
states, however the entanglement of the clones is reduced. The most entangled are clones 
obtained using the cloning machine, which is specially designed to preserve 
the entanglement, but which is rather poor as the cloner. 
Conversely, the universal cloning machine clones states quite well, 
but the clones obtained are almost completely devoid of entanglement. 

The results obtained seems to suggest that the quantum cloning cannot actually be used to improve
the performance of the quantum computation tasks due to the drastic reduction of the entanglement 
in the cloning process.


\begin{thebibliography}{00}
\bibitem{Zurek} W. K. Wootters and W. H. Zurek, 1982, {\em Nature (London)}, {\bf 299}, 802.
\bibitem{Buzek_asym} M. Hillery and V. Bu\v{z}ek, 1997, {\em Phys. Rev. A}, {\bf 56}, 1212. 
%\bibitem{Bruss_asym} D. Bru\ss, D. P. DiVincenzo, A. Ekert, Ch. A. Fuchs, Ch. Macchiavello, 
%							and J. A. Smolin, 1998, {\em Phys. Rev. A}, {\bf 57}, 2368.
\bibitem{Bruss_asym} D. Bru\ss \hspace{0.08cm} {\em et al.}, 1998, {\em Phys. Rev. A}, {\bf 57}, 2368.
\bibitem{Barnett_asym} A. Chefles and S. M. Barnett, 1999, {\em Phys. Rev. A}, {\bf 60}, 136.
\bibitem{Buzek_beyond} V. Bu\v{z}ek and M. Hillery, 1996, {\em Phys. Rev. A}, {\bf 54}, 1844. 
\bibitem{Buzek_network} V. Bu\v{z}ek, S. L. Braunstein, M. Hillery, and D. Bru\ss, 1997, {\em Phys. Rev. A}, {\bf 56}, 3446.
\bibitem{Werner_d1} R. F. Werner, 1998, {\em Phys. Rev A}, {\bf 58}, 1827.
\bibitem{Buzek_d} V. Bu\v{z}ek and M. Hillery, 1998, {\em Phys. Rev. Lett.}, {\bf 81}, 5003.
\bibitem{Werner_d2} M. Keyl and R. F. Werner, 1999, {\em J. Math. Phys.}, {\bf 40}, 3283.
\bibitem{Hardy_qcomp} E. F. Galv\~{a}o and L. Hardy, 2000, {\em Phys. Rev. A}, {\bf 62}, 022301. 
\bibitem{Buzek_class}R. Derka, V. Buzek, and A. K. Ekert, 1998, {\em Phys. Rev. Lett.}, {\bf 80}, 1571.
\bibitem{Masiak} P. Masiak and P. L. Knight, 2001, {\em Fortschr. Phys.}, {\bf 49}, 1001.
\bibitem{Ceft} N. J. Ceft, 1998, {\em Acta Phys. Slov.}, {\bf 48}, 115.
\bibitem{Aspect1} A. Aspect, P. Grangier, and G. Roger, 1981, {\em Phys.Rev.Lett.}, {\bf 47}, 460.
\bibitem{Aspect2} A. Aspect, P. Grangier, and G. Roger, 1982, {\em Phys.Rev.Lett.}, {\bf 49}, 91.
\bibitem{Vedral_ent1} V. Vedral, M. B. Plenio, M. A. Rippin, and P. L. Knight, 1997, {\em Phys. Rev. Lett.}, {\bf 78}, 2275.
\bibitem{Vedral_ent2} V. Vedral, M. B. Plenio, K. Jacobs, and P. L. Knight, 1997, {\em Phys. Rev. A}, {\bf 56}, 4452.
\bibitem{Wootters1} S. Hill and W. K. Wootters, 1997, {\em Phys. Rev. Lett.}, {\bf 78}, 5022.
\bibitem{Wootters2} W. K. Wootters, 1998, {\em Phys. Rev. Lett.}, {\bf 80}, 2245.
\end{thebibliography}
\end{document}